\documentclass[prc,letterpaper,twocolumn,showpacs,showkeys,lengthcheck,
               floatfix,nofootinbib,preprintnumbers,superscriptaddress]{revtex4-1} 

\usepackage{mathtools}
\usepackage{amssymb,amsfonts}
\usepackage{bm}
\usepackage{slashed}

\usepackage[svgnames]{xcolor}
\usepackage[english]{babel}
\usepackage{blindtext}
\usepackage{microtype}
\usepackage{tikz}
\usepackage{dcolumn}
\usepackage{multirow}
\usepackage[]{units}

\usepackage{graphicx}
\usepackage[caption=false]{subfig}

\usepackage[colorlinks=true]{hyperref}
\usepackage{ifpdf}

\def\a{\alpha}
\def\b{\beta}

\def\g{\gamma}

\def\ve{\varepsilon}

\def\l{\lambda}
\def\s{\sigma}

\def\G{\Gamma}
\def\L{\Lambda}

\def\S{\Sigma}

\def\hs{\hspace}

\def\no{\nonumber}

\def\lf{\left}
\def\rg{\right}

\newcommand{\sh}[1]{\slashed{#1}}

\font\bb=bbmss10 scaled 1200
\def\ident{\mbox{\bb 1}}

\def\be{\begin{equation}}
\def\ee{\end{equation}}
\def\nn{\nonumber}

\begin{document}

%\title{Electromagnetic Form Factors of the Baryon Octet in the NJL model}
%\title{Electromagnetic Form Factors of the Baryon Octet in a confining NJL model}
\title{Baryon Octet Electromagnetic Form Factors in a confining NJL model}

\author{Manuel~E.~Carrillo-Serrano}
\affiliation{CSSM and ARC Centre of Excellence for Particle Physics at the Tera-scale,\\
Department of Physics,
University of Adelaide, Adelaide SA 5005, Australia
}

\author{Wolfgang Bentz}
\affiliation{Department of Physics, School of Science, Tokai University,
Hiratsuka-shi, Kanagawa 259-1292, Japan}

\author{Ian~C.~Clo\"et}
\affiliation{Physics Division, Argonne National Laboratory, Argonne, Illinois 60439, USA
}

\author{Anthony~W.~Thomas}
\affiliation{CSSM and ARC Centre of Excellence for Particle Physics at the Tera-scale,\\
Department of Physics,
University of Adelaide, Adelaide SA 5005, Australia
}

\begin{abstract}
Electromagnetic form factors of the baryon octet are studied using a Nambu--Jona-Lasinio model 
which utilizes the proper-time regularization scheme to simulate aspects of colour confinement. 
In addition, the model also incorporates corrections to the dressed quarks from vector meson 
correlations in the $t$-channel and the pion cloud. Comparison with recent chiral extrapolations
of lattice QCD results shows a remarkable level of consistency. For the charge radii we find the 
surprising result that $r_E^p < r_E^{\Sigma^+}$ and $\lf|r_E^n\rg| < |r_E^{\Xi^0}|$,
whereas the magnetic radii have a pattern largely consistent with a naive expectation based
on the dressed quark masses.
\end{abstract}


\maketitle
%===============================================================================
%===============================================================================
\section{Introduction}

The lowest mass baryon octet plays a special role in 
the quest to understand the strong 
interaction. Along with their masses and axial charges, it is particularly important 
to explain their distributions of charge and magnetisation in terms of the underlying 
quark-gluon dynamics. Empirically, these distributions are expressed by 
their electromagnetic form factors, which present an extraordinary challenge for
QCD~\cite{Arrington:2006zm}. Considerable experimental 
effort has been devoted to the measurement and parametrization of the 
electromagnetic form factors of the 
nucleon~\cite{Brodsky:1973kr,Brodsky:1974vy,Farrar:1979aw,Tadevosyan:2007yd,Jones:1999rz,Gayou:2001qt,Gayou:2001qd,Ralston:2003mt,Kelly:2004hm,Perdrisat:2006hj,Bernauer:2010wm,Puckett:2011xg}.
However, for the other members of the baryon octet this 
is a more difficult task because of their short lifetimes~\cite{Agashe:2014kda}. 

Theoretical predictions for nucleon electromagnetic form factors, and for example,
parton distribution functions, have been made using a variety of approaches, such as quark models~\cite{LeYaouanc:1976ne,Isgur:1979be,Thomas:1981vc,Diakonov:1996sr,Diakonov:1997sj,Mineo:1999eq,Cloet:2005pp,Cloet:2005rt,Cloet:2007em,Bentz:2007zs,Matevosyan:2011vj,Cloet:2014rja}, QCD sum rules~\cite{Aliev:2013jda}, the Dyson-Schwinger equations~\cite{Cloet:2008re,Eichmann:2011vu,Segovia:2014aza} and lattice QCD
simulations~\cite{Lin:2007ap,Durr:2008zz,WalkerLoud:2008bp,Aoki:2008sm,Shanahan:2012wa,Shanahan:2013apa,Borsanyi:2014jba,Shanahan:2014cga,Shanahan:2014uka}.
For the other elements of the octet, prior to lattice QCD computations, early work on the
 spectrum, electromagnetic form factors and weak form factors was based on, for example, the 
bag model~\cite{Chodos:1974pn,Theberge:1980ye,Theberge:1981pu,Myhrer:1982sp,Kubodera:1984qd,Tsushima:1988xv,Yamaguchi:1989sx,Wagner:1998fi,Boros:1999tb,Thomas:1999mu,Thomas:2000fa,Bass:2009ed},
QCD sum rules~\cite{Aliev:2013jda}, constituent quark models~\cite{Silva:2005vp,Ramalho:2012pu,Carrillo-Serrano:2014zta} 
and more recently the Dyson-Schwinger equations~\cite{Sanchis-Alepuz:2015fcg}.

With the advent of more precise lattice QCD computations, together
with chiral extrapolations to the physical point, 
the baryon octet spectrum has been accurately 
reproduced~\cite{Durr:2008zz,Shanahan:2012wa,Young:2009zb} 
and more recently the electromagnetic form factors of the outer ring of the octet have been
extracted~\cite{Shanahan:2014cga,Shanahan:2014uka}.
At the same time, the recent work in Ref.~\cite{Cloet:2014rja} showed
promising results when the Nambu--Jona-Lasinio (NJL) 
model~\cite{Nambu:1961tp,Nambu:1961fr} was applied to 
the calculation of the nucleon electromagnetic form factors~\cite{Cloet:2014rja}. 
In addition, the model has also been applied to 
the axial charges in several $\Delta S = 0$ $\b$-decays in the 
baryon octet~\cite{Carrillo-Serrano:2014zta}, 
and the electromagnetic form factors of 
the $\rho$ meson~\cite{Carrillo-Serrano:2015uca}.

In the present work we extend the framework developed in Ref.~\cite{Cloet:2014rja} for the nucleon, to a description of the electromagnetic form factors of the baryon octet. 

%===============================================================================
%===============================================================================
\section{Baryons in a Nambu--Jona-Lasinio Model\label{sec:NJL}}
Extensive reviews of the NJL model exist~\cite{Klevansky:1992qe,Hatsuda:1994pi,Vogl:1991qt} and here we use the $SU(3)$ flavour
version with only the four-fermion interaction. The Lagrangian in the $\bar{q}q$ interaction channel, 
which we take in Fierz symmetric form~\cite{Klevansky:1992qe}, reads
\begin{align}
\mathcal{L} &= \bar{\psi}\lf(i\sh{\partial} - \hat{m}\rg)\psi 
+ \tfrac{1}{2}\,G_\pi\Big[\lf(\bar{\psi}\,\l_i\,\psi\rg)^2 
- \lf(\bar{\psi}\,\g_5\,\l_i\,\psi\rg)^2\Big] \no \\
&- \tfrac{1}{2}\,G_\rho\lf[\lf(\bar{\psi}\,\g^\mu\,\l_i\,\psi\rg)^2 
+ \lf(\bar{\psi}\,\g^\mu\g_5\,\l_i\,\psi\rg)^2\rg], 
\label{eq:lagrangian}
\end{align}
with $\hat{m} = \text{diag}\lf[m_u,\,m_d,\,m_s\rg]$ and $\l_i$ 
the eight Gell-Mann matrices plus $\l_0 \equiv \sqrt{2/3}\,\ident$. 
Gluon degrees of freedom are absent in the NJL model and therefore one must specify 
a method of regularization. We use the proper-time scheme, because it simulates aspects of quark
confinement~\cite{Ebert:1996vx,Hellstern:1997nv,Bentz:2001vc}.

The dressed quark mass, $M_q$, of flavour $q=u,d,s$ is obtained by 
solving the gap equation. With proper-time regularization
$M_q$ satisfies~\cite{Carrillo-Serrano:2014zta}
\begin{align}
M_q = m_q + \frac{3}{\pi^{2}}\,M_q\,G_{\pi}
\int_{1/\L_{UV}^{2}}^{1/\L_{IR}^{2}} d\tau\
\frac{e^{-\tau M_q^2}}{\tau^2}.
\label{gap}
\end{align}
We note that in the $SU(3)$ flavour case, flavour mixing is absent,
in contrast to the $SU(2)$ flavour case~\cite{Klevansky:1992qe}.

When solving the 3-body problem~\cite{Afnan:1977pi} in the NJL model,
to obtain the Faddeev vertex functions for each member of the baryon 
octet, strong diquark correlations naturally appear.
To determine the diquark $t$-matrices it is therefore convenient to make 
a different Fierz transformation on the $SU(3)$ NJL Lagrangian density,
which yields the effective $qq$ interactions~\cite{Ishii:1995bu} in 
the scalar and axial-vector diquark channels:
\begin{align}
\mathcal{L}_{I}^{qq} &= G_s \Bigl[\bar{\psi}\,\g_5\, C\,\l_a\,\beta_A\, \bar{\psi}^T\Bigr]
\Bigl[\psi^T\,C^{-1}\g_5\,\l_a\,\beta_A\, \psi\Bigr] \no \\
&
+ G_a \Bigl[\bar{\psi}\,\g_\mu\,C\,\l_s\,\beta_A\, \bar{\psi}^T\Bigr]
\Bigl[\psi^T\,C^{-1}\g^{\mu}\,\l_s\, \beta_A\, \psi\Bigr] \, ,
\label{eq:qqlagrangian}
\end{align}
where $C = i\g_2\g_0$ is the charge conjugation matrix, and $G_s$ and $G_a$ are the 
couplings in the scalar and axial-vector diquark channels, respectively. The quark flavour matrices 
are represented by $\l_a$ for $a \in (2,5,7)$ and $\l_s$ for $s \in (0,1,3,4,6,8)$, while 
$\beta_A = \sqrt{3/2}\,\lambda_A~(A=2,5,7)$ selects the colour
$\bar{3}$ states~\cite{Ishii:1995bu,Ishii:1993np,Ishii:1993rt}. 

%===============================================================================
\begin{figure}[tbp]
\centering\includegraphics[width=\columnwidth,clip=true,angle=0]{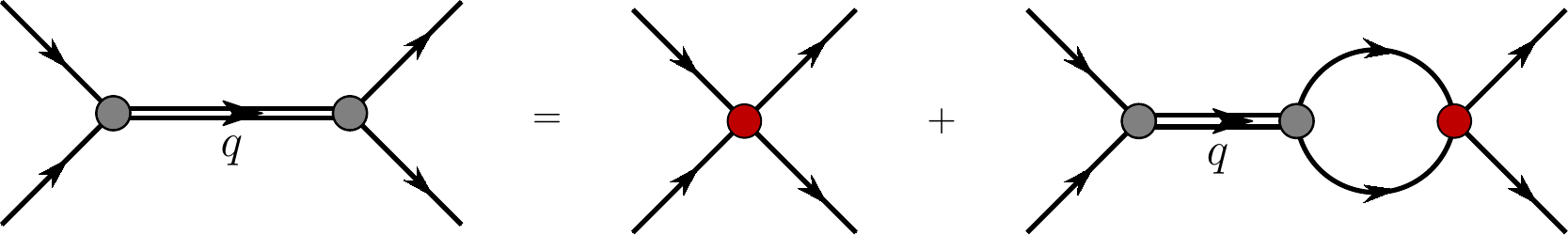}
\caption{(Colour online) Inhomogeneous Bethe-Salpeter equation for quark--quark (diquark) correlations.}
\label{fig:1}
\end{figure}
%===============================================================================

Fig.~\ref{fig:1} depicts the Bethe-Salpeter equation (BSE) which describes two-particle 
($qq$ in this case) bound states. Solutions to the BSE for the scalar and axial-vector 
diquarks, in terms of the reduced $t$-matrices, are expressed as
\begin{align}
\label{eq:tscalar}
\tau_{[q_1q_2]}(q) 
&= \frac{4i\,G_s}{1 + 2\,G_s\,\Pi_{[q_1q_2]}(q^2)}, \\[0.0ex]
\label{eq:taxial}
\tau^{\mu\nu}_{\{q_1q_2\}}(q) 
&= \frac{4\,i\,G_a}{1 + 2\,G_a\,\Pi^T_{\{q_1 q_2\}}(q^2)}\lf(g^{\mu\nu} - \frac{q^\mu q^\nu}{q^2}\rg) \no \\
&\hs{0mm}
+ \frac{4\,i\,G_a}{1 + 2\,G_a\,\Pi^L_{\{q_1 q_2\}}(q^2)}\,\frac{q^\mu q^\nu}{q^2}.
\end{align}
The bubble diagrams are given by
\begin{align}
\label{eq:bubble_PP}
&\Pi_{[q_1q_2]}\lf(q^2\rg) = 6i \int \frac{d^4k}{(2\pi)^4}\ \mathrm{Tr}\lf[\g_5\,S_{q_1}(k)\,\g_5\,S_{q_2}(k+q)\rg], \\
\label{eq:bubble_VV}
&\Pi^T_{\{q_1q_2\}}(q^2)\lf(g^{\mu\nu} - \frac{q^\mu q^\nu}{q^2}\rg) + \Pi^L_{\{q_1q_2\}}\frac{q^\mu q^\nu}{q^2} = \no \\
&\hs{17mm}6i \int \frac{d^4k}{(2\pi)^4}\ \mathrm{Tr}\lf[\g^\mu\,S_{q_1}(k)\,\g^\nu\,S_{q_2}(k+q)\rg],
\end{align}
where $S_q(k) = [\sh{k} - M_q + i\ve]^{-1}$ is the dressed quark propagator and the trace is over Dirac indices only.
Throughout this paper square brackets will represent a scalar diquark and 
curly brackets an axial-vector diquark, where $q_1$ and $q_2$ label the flavour $(u,d,s)$
of each quark inside the diquark. In the solution to the Faddeev equation we employ the pole approximation
for the reduced $t$-matrices~\cite{Cloet:2014rja,Carrillo-Serrano:2014zta}:
\begin{align}
\label{eq:scalarpropagatorpoleform}
\tau_{[q_1q_2]}(q) &\to  \frac{-i\,Z_{[q_1q_2]}}{q^2 - M_{[q_1q_2]}^2+ i\ve} \, , \\
\label{eq:axialpropagatorpoleform}
\tau^{\mu\nu}_{\{q_1q_2\}}(q) &\to \frac{-i\,Z_{\{q_1q_2\}}}{q^2 - M_{\{q_1q_2\}}^2 + i\ve} 
\lf(g^{\mu\nu} - \frac{q^{\mu}q^{\nu}}{M_{\{q_1q_2\}}^2}\rg) \, .
\end{align}
where the $Z$'s are the residues at the poles~\cite{Cloet:2014rja,Carrillo-Serrano:2014zta}.

Solutions to the Faddeev equation for each member of the baryon octet in this model have already
been detailed in Ref.~\cite{Carrillo-Serrano:2014zta}, therefore here we will just 
give a brief review. For each baryon the Faddeev equation, in the static approximation~\cite{Buck:1992wz},
takes the general form
\begin{align}
\G_{b}(p,s) = Z_b\ \Pi_b(p)\ \G_{b}(p,s),
\label{eq:faddeev}
\end{align}
where $b = N, \Sigma, \Xi$ labels the baryon and the $p^2$ that satisfies this equation
defines the baryon mass. The quark exchange kernel is labelled by $Z_b$ and $\Pi_b(p)$
contains the quark--diquark bubble diagrams. The Faddeev vertex is normalized such that $\G_{b}(p,s) = \sqrt{-\mathcal{Z}_b}\ \G_{0b}(p,s)$, 
where $\mathcal{Z}_b$ is given by
\begin{align}
\mathcal{Z}_b^{-1} = \lf.\overline{\G}_{0b}\ 
\frac{\partial\,\Pi_b(p)}{\partial \sh{p}}\ \G_{0b}\rg|_{p^2 = M_b^2}.
\label{eq:z_b}
\end{align}
We normalize the vertex $\G_{0b}(p,s)$ such that $\overline{\G}_{0b}\,\G_{0b} = 1$.
%
%===============================================================================
\begin{figure}[tbp]
\centering\includegraphics[width=\columnwidth,clip=true,angle=0]{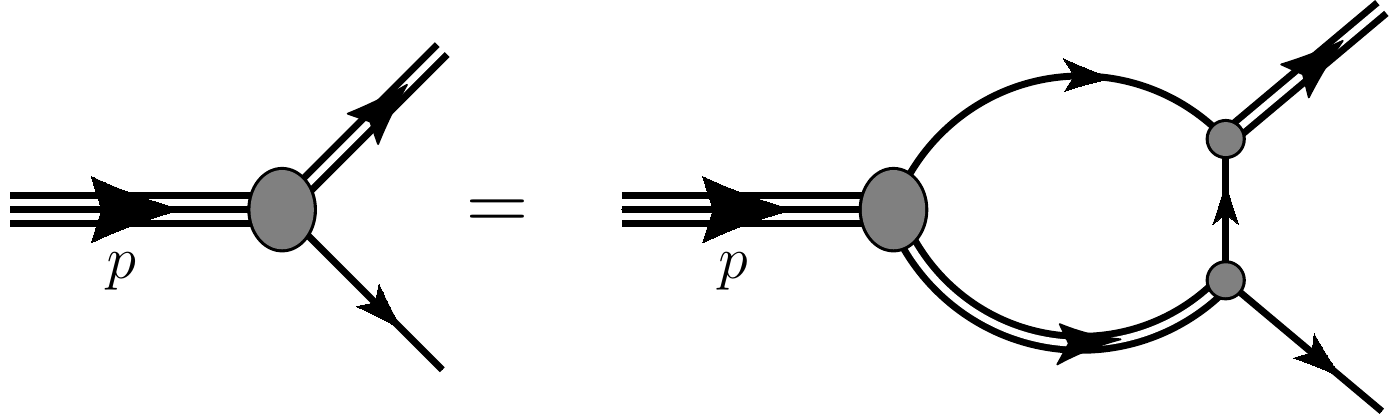}
\caption{Homogeneous Poincar\'e covariant Faddeev equation whose 
solution gives the mass and 
vertex function for each member of the baryon octet.}
\label{fig:2}
\end{figure}
%===============================================================================

For the form factor calculations we will only consider the nucleon, $\Sigma^{\pm}$ and $\Xi$,
as there are no lattice results for the $\L$ and the $\Sigma^0$.
The Faddeev vertex functions are evaluated for equal light quark masses 
($M_u = M_d \equiv  M_\ell$) and for the members of the baryon octet considered, the 
Dirac structure is
\begin{align}
\G_b(p,s) &= \begin{bmatrix}
\G_{q_1[q_1q_2]}(p,s) \\[1.1ex]
\G_{q_1\{q_1q_2\}}^\mu(p,s) \\[1.1ex]
\G_{q_2\{q_1q_1\}}^\mu(p,s)
\end{bmatrix}, \no \\
&= \sqrt{-\mathcal{Z}_b}\,\begin{bmatrix} 
\a_1 \\[0.8ex]
\a_2\,\frac{p^\mu}{M_b}\,\g_5 + \a_3\,\g^\mu\g_5 \\[0.8ex]
\a_4\,\frac{p^\mu}{M_b}\,\g_5 + \a_5\,\g^\mu\g_5 
\end{bmatrix}u_b(p,s) \, .  
\label{eq:vertex}
\end{align}
The quark exchange kernel reads
\begin{align}
Z_b &= 3
\begin{bmatrix} 
 \frac{1}{M_{q_1}}                  & \frac{1}{M_{q_1}}\g_\s\g_5         & -\frac{\sqrt{2}}{M_{q_2}}\g_\s\g_5 \\[0.8ex]
 \frac{1}{M_{q_1}}\g_5\g_\mu         & \frac{1}{M_{q_1}}\g_\s\g_\mu        & \frac{\sqrt{2}}{M_{q_2}}\g_\s\g_\mu \\[0.8ex]
-\frac{\sqrt{2}}{M_{q_2}}\g_5\g_\mu  & \frac{\sqrt{2}}{M_{q_2}}\g_\s\g_\mu & 0
\end{bmatrix},
\label{eq:kernel}
\end{align}
where, following Ref.~\cite{Carrillo-Serrano:2014zta}, 
$M_{q_1}$ and $M_{q_2}$ correspond to
the masses of the singly and doubly represented quark, respectively. 
Projecting the Faddeev kernel onto a colour singlet
gives the factor of $3$ in Eq.~\eqref{eq:kernel}.

%===============================================================================
%===============================================================================
%\section{Parameter determination\label{sec:Parameters}}
%
%===============================================================================
\begin{table}[tbp]
\addtolength{\tabcolsep}{5.0pt}
\addtolength{\extrarowheight}{2.2pt}
\begin{tabular}{ccccccccc}
\hline\hline
$\L_{IR}$ & $\L_{UV}$ & $M_\ell$  & $M_s$  & $G_\pi$ & $G_\rho$ & $G_s$  & $G_a$ \\
\hline
0.240    & 0.645    & 0.40     & 0.56   & 19.0   & 11.0     & 5.8    & 4.9   \\
\hline\hline
\end{tabular}
\caption{Model parameters, where all masses and regularization parameters are given in units of GeV, 
while the Lagrangian couplings are in units of GeV$^{-2}$.} 
\label{tab:parameters}
\end{table}
%===============================================================================
%
%===============================================================================
\begin{table}[tbp]
\addtolength{\tabcolsep}{-0.5pt}
\addtolength{\extrarowheight}{2.2pt}
\begin{tabular}{cccccccccc}
\hline\hline
$M_{[\ell\ell]}$ & $M_{[\ell s]}$ & $M_{\{\ell\ell\}}$ & $M_{\{\ell s\}}$ & $M_{\{ss\}}$ & $Z_{[\ell\ell]}$ & $Z_{[\ell s]}$ & $Z_{\{\ell\ell\}}$ & $Z_{\{\ell s\}}$ & $Z_{\{ss\}}$ \\[0.2em]
\hline
0.768 & 0.903 & 0.929 & 1.04 & 1.15 & 11.1   & 12.0  & 6.73          & 7.54         & 8.36           \\
\hline\hline
\end{tabular}
\caption{Results for the diquark masses and pole residues in 
the various diquark $t$-matrices
[c.f. Eqs.~\eqref{eq:scalarpropagatorpoleform} and 
\eqref{eq:axialpropagatorpoleform}]. 
All masses are in GeV and the residues are dimensionless.}
\label{tab:massesresidues}
\end{table}
%===============================================================================

%===============================================================================
\begin{table}[t]
\addtolength{\tabcolsep}{9.5pt}
\addtolength{\extrarowheight}{2.2pt}
\begin{tabular}{c|cccc}
\hline\hline
      & $M_N$ & $M_\L$  & $M_\Sigma$ & $M_\Xi$    \\
\hline
NJL   & 0.940 & 1.126  & 1.170    & 1.277    \\
Experiment & 0.940 & 1.116  & 1.193    & 1.318    \\
\hline\hline
\end{tabular}
\caption{Calculated octet baryon masses are compared with the average experimental 
mass for the corresponding multiplet. 
Note that the nucleon mass was used to constrain an NJL model parameter. 
All masses are in units of GeV.} 
\label{tab:octetmasses}
\end{table}
%===============================================================================

%===============================================================================
\begin{table}[t]
\addtolength{\tabcolsep}{3.5pt}
\addtolength{\extrarowheight}{2.2pt}
\begin{tabular}{c|cccccccc}
\hline\hline
        & $\a_1$ & $\a_2$ & $\a_3$ & $\a_4$ & $\a_5$ & $\mathcal{Z}_{B}$  \\
\hline
nucleon & 0.552 & 0.031 &-0.233 &-0.043 & 0.329 & 28.136 \\
%$\L$    & 0.481 & 0.328 &-0.048 & 0.394 & --    & 19.482 \\
$\S$    & 0.506 & 0.066 &-0.211 &-0.051 & 0.352 & 20.041 \\
$\Xi$   & 0.525 & 0.046 &-0.249 &-0.044 & 0.324 & 18.819 \\
\hline\hline
\end{tabular}
\caption{Coefficients that define the Faddeev vertex functions for 
each member of the baryon octet considered herein.} 
\label{tab:vertex}
\end{table}
%===============================================================================

The parameters employed here are summarized in Tab.~\ref{tab:parameters}.
The infrared cutoff should be of the order of $\L_{\text{QCD}}$ because it 
implements quark confinement~\cite{Hellstern:1997nv,Bentz:2001vc}, and we choose 
$\L_{IR} = 0.240\,$GeV.
The masses of the light dressed quarks are chosen as $M_u=M_d=M_\ell=0.4\,$GeV, 
while the $s$-quark mass, $M_s$, is chosen to reproduce the  
mass of the $\Omega^-$ baryon.
The parameters $\L_{UV}$, $G_{\pi}$ and $G_{\rho}$ are fit to reproduce 
the empirical values of the pion decay
constant, and the pion and $\rho$ masses, while $G_a$ and $G_s$ are fixed by the 
physical $\Delta^{++}$ and nucleon masses. 

In Tab.~\ref{tab:massesresidues} we summarize 
the results for the diquark masses, as well as the residues for the diquark $t$-matrices 
given in Eqs.~\eqref{eq:scalarpropagatorpoleform} and~\eqref{eq:axialpropagatorpoleform}.
The octet baryon masses, obtained by solving the appropriate Faddeev 
equation, are given in Tab.~\ref{tab:octetmasses}. 
The parameters defining the Faddeev vertex function for each member of 
the baryon octet are summarized in Tab.~\ref{tab:vertex}.

%===============================================================================
%===============================================================================
\section{Baryon form factors\label{sec:FF}}
\begin{figure}[tbp]
\centering\includegraphics[width=\columnwidth,clip=true,angle=0]{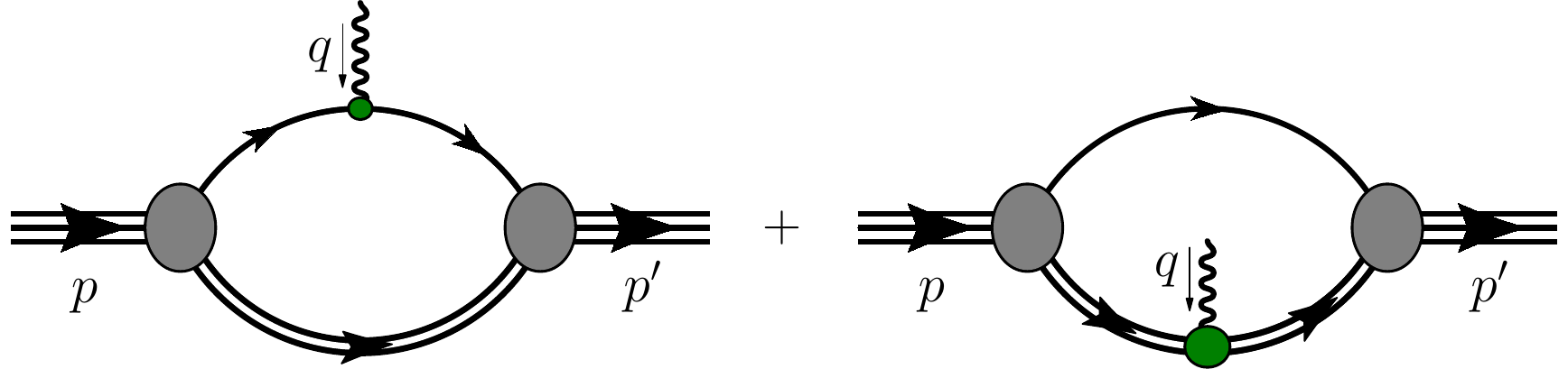}
\caption{(Colour online) Feynman diagrams representing the electromagnetic 
current for 
the octet baryons. The diagram on the left is called the 
``quark diagram'' and the one on 
the right the ``diquark diagram''. 
In the diquark diagram the photon interacts with each 
quark inside the diquark.}
\label{fig:nucleon_current}
\end{figure}

The electromagnetic form factors, $F_{1b}$ and $F_{2b}$, of an octet baryon $b$,  
are defined by the electromagnetic current 
\begin{align}
\label{Eq:EmCurrentBaryon}
&j^{\mu,b}_{\l'\l}(p',p) = \lf<p',\l'\lf|J_{em}^{\mu}\rg|p,\l\rg>, \\
&= \bar{u}_b(p',\l')\left[\g^\mu\, F_{1b}(Q^2) + 
\frac{i\sigma^{\mu\nu}q_{\nu}}{2\,M_{b}}\,F_{2b}(Q^2)\right]u_b(p,\l), \nn
\end{align}
where $\l$ and $\l'$ represent the helicity of the incoming and outgoing baryon and
$q$ is the 4-momentum transfer, where $Q^2 = -q^2$. In the NJL model considered here,
this electromagnetic current is represented by the Feynman diagrams illustrated in Fig.~\ref{fig:nucleon_current}.

In the evaluation of the baryon form factors we dress the quark-photon 
vertices by including both vector meson correlations in the $t$-channel, 
through the inhomogeneous BSE, and also effects 
from pion loops. This formalism is described in detail in Ref.~\cite{Cloet:2014rja}. 
In summary, the dressed quark-photon vertex has the form:
\begin{align}
\L^\mu_{\g Q}(p',p)=\g^\mu\,F_{1Q}(Q^2) + \frac{i\s^{\mu\nu}q_\nu}{2M_q}\,F_{2Q}(Q^2),
\label{Eq:QuarkVertex}
\end{align}
where $Q = (U,D,S)$ and the $F_{2Q}$ form factor results from the pion loop corrections.
Explicit expressions for the dressed quark form factors can be found in Ref.~\cite{Cloet:2014rja}, 
supplemented here with the dressed strange quark form factors: $F_{1S} = -\tfrac{1}{3}F_{1\phi}$ 
and $F_{2S} = 0$, where $F_{1\phi}$ is generated by $t$-channel $\phi$ meson correlations~\cite{Cloet:2014rja}. 
Because the $\pi$ is much lighter than the $K$ we expect pion loops to give the dominant chiral correction, and therefore 
omit $K$ loops~\cite{Ninomiya:2014kja}.

%===============================================================================
%===============================================================================

The total baryon form factors have the form
\begin{align}
F_{ib}(Q^2) = \sum_{Q}\left[F_{1Q}\,f_{ib}^{Q,V} + F_{2Q}\,f_{ib}^{Q,T}\right],
\label{eq:tot_b_ff}
\end{align}
where the sum is over the dressed quarks in each baryon and 
the $Q^2$ dependence for each form factor is implicit.
The body form factors, $f_{ib}^{Q,V}$ and $f_{ib}^{Q,T}$, are
given by the Feynman diagrams of Fig.~\ref{fig:nucleon_current}, where the former
is obtained from a point-like quark-photon vector coupling $\g^{\mu}$ and the latter 
a point-like quark-photon tensor coupling $i\s^{\mu\nu}q_{\nu}/2M_q$. Each of these
body form factors contains contributions from both the quark and diquark diagrams
illustrated in Fig.~\ref{fig:nucleon_current}, and further details can be found in
Refs.~\cite{Cloet:2014rja,Carrillo-Serrano:2014zta}

The Sachs form factors are defined by
\begin{align}
G_{Eb} = F_{1b} - \frac{Q^2}{4M_{b}^2}\,F_{2b}, \qquad G_{Mb} = F_{1b} + F_{2b},
\label{eq:GEGM}
\end{align}
and in Tab.~\ref{Tab:EMFF} results for $G_{Mb}$ at $Q^2=0$ are given, 
with both the vector meson dressing of the quark-photon vertices (labelled BSE)
and also with the effect of the pion cloud as well. 
It is evident that the effect of the pion cloud is to 
increase the magnitude of the magnetic moments across the octet, 
almost uniformly improving agreement with experiment. The exception 
is the $\Sigma^-$, where the discrepancy is about 40\%, 
which suggests that in this case the effect of the pion cloud may be 
overestimated~\cite{Thomas:1999mu,Thomas:2000fa}.

%===============================================================================
\begin{table}[tbp]
\addtolength{\tabcolsep}{13.0pt}
\addtolength{\extrarowheight}{2.2pt}
\begin{tabular}{c|ccc}
\hline\hline
 &  $\mu^{(BSE)}_b$ & $\mu_b$ & $\mu_b^{exp}$\\
\hline
proton   &  2.43 &  2.78 & 2.793 \\
neutron  & -1.25 & -1.81 &-1.913 \\
$\S^+$   &  2.30 &   2.62 & 2.458(10) \\
$\S^-$   & -1.04 & -1.62 &-1.160(25) \\
$\Xi^0$  & -1.08 & -1.14 &-1.250(14) \\
$\Xi^-$  & -0.78 & -0.67 &-0.6507(25) \\
\hline\hline
\end{tabular}
\caption{Magnetic moments in units of nuclear magnetons. The BSE results include only the
vector meson contributions to the dressed quark form factors, while the final
results also include effects from the pion cloud. 
A comparison with the experimental values~\cite{Agashe:2014kda} is shown.} 
\label{Tab:EMFF}
\end{table}
%===============================================================================

Results for the charge and  magnetic radii of the octet baryons, defined with respect 
to the Sachs form factors, are summarized in Tab.~\ref{Tab:Radii}. The PL column
stands for a structureless, point-like quark and the other two columns 
use either the BSE or the fully dressed quark-photon vertex.
The effect of the vector meson and pion cloud dressing on these quantities is evident. 
In all cases the radii increase with the inclusion of vector meson corrections, sometimes
dramatically. The effect of the pion cloud alone is also 
to increase the radii, except for the $\Xi^-$, where $r_M^{\Xi^-}$ is
reduced from 0.62\,fm with only the BSE vertex to 0.51\,fm including the pion cloud. 
The maximum contribution appears in the neutron charge radius 
where we find an increase of 85\% from the pion cloud alone. 
The smallest contribution of the pion cloud occurs for the magnetic radius of
the $\Xi^0$ and $\Xi^-$, which is to be expected because the strange quarks do not
couple to the pion cloud.

Surprisingly, we find that $r_E^p < r_E^{\Sigma^+}$ and $\lf|r_E^n\rg| < |r_E^{\Xi^0}|$. This is 
because the difference between $m_{\rho(\omega)}$ and $m_{\phi}$ (about 250 MeV), which characterize 
the vector meson dressing of the quark-photon vertices, makes the slope of $F_{1D}$ at $Q^2=0$ around $1.78$ times larger than
that for $F_{1S}$ in the BSE case, and $2.45$ times larger including the pion cloud. The contributions from these form factors tend to lower the
charge radius, suppressing $r_E$ more in the nucleon than in the $\Sigma^+$ or $\Xi^0$.
In addition, for the proton the term arising from the $F_{2D}$ form factor reduces 
the proton radius even further, and this term is absent in the $\Sigma^+$
because the strange quark does not couple to the pion field.

%===============================================================================
\begin{figure*}[tbp]
  \subfloat{\centering\includegraphics[width=1.0\columnwidth,clip=true,angle=0]{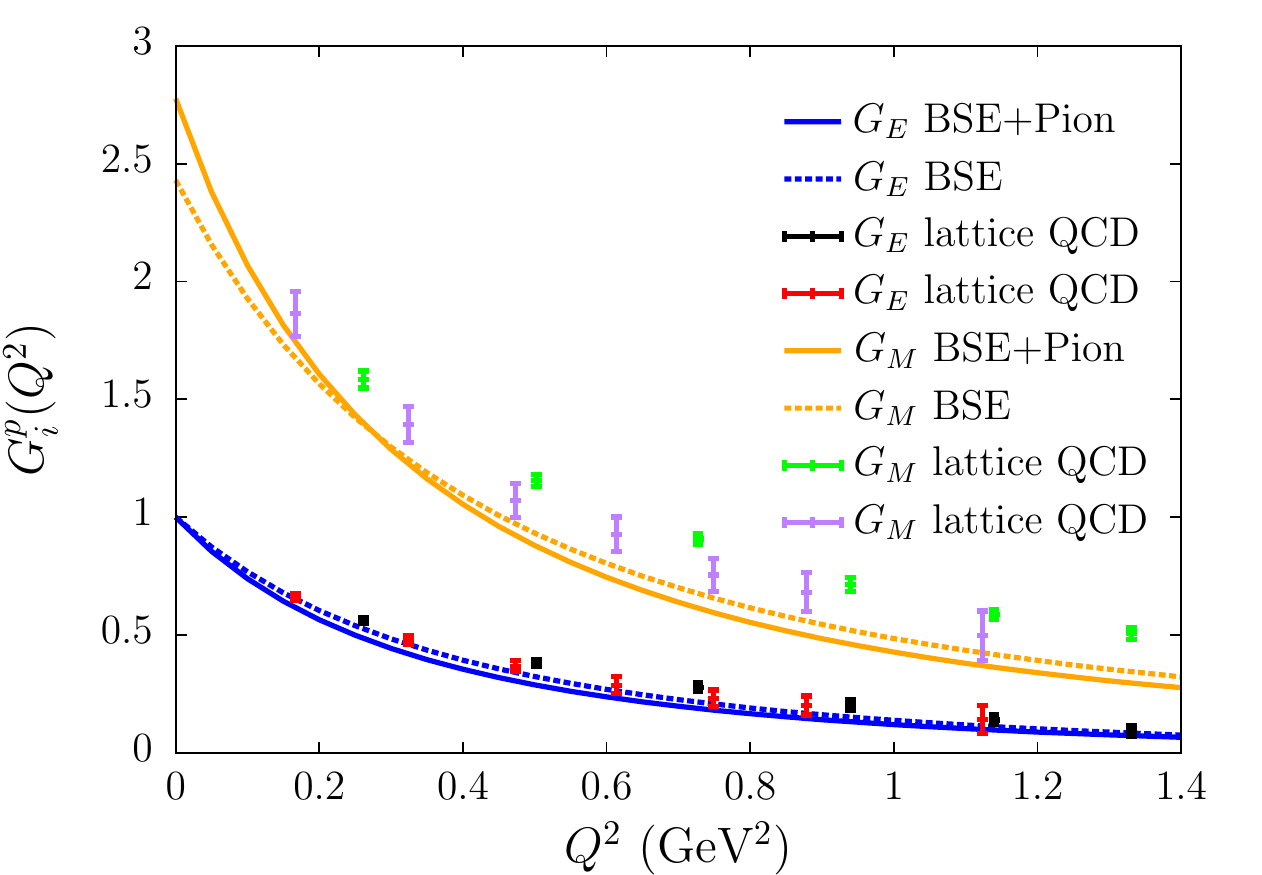}}
  \subfloat{\centering\includegraphics[width=1.0\columnwidth,clip=true,angle=0]{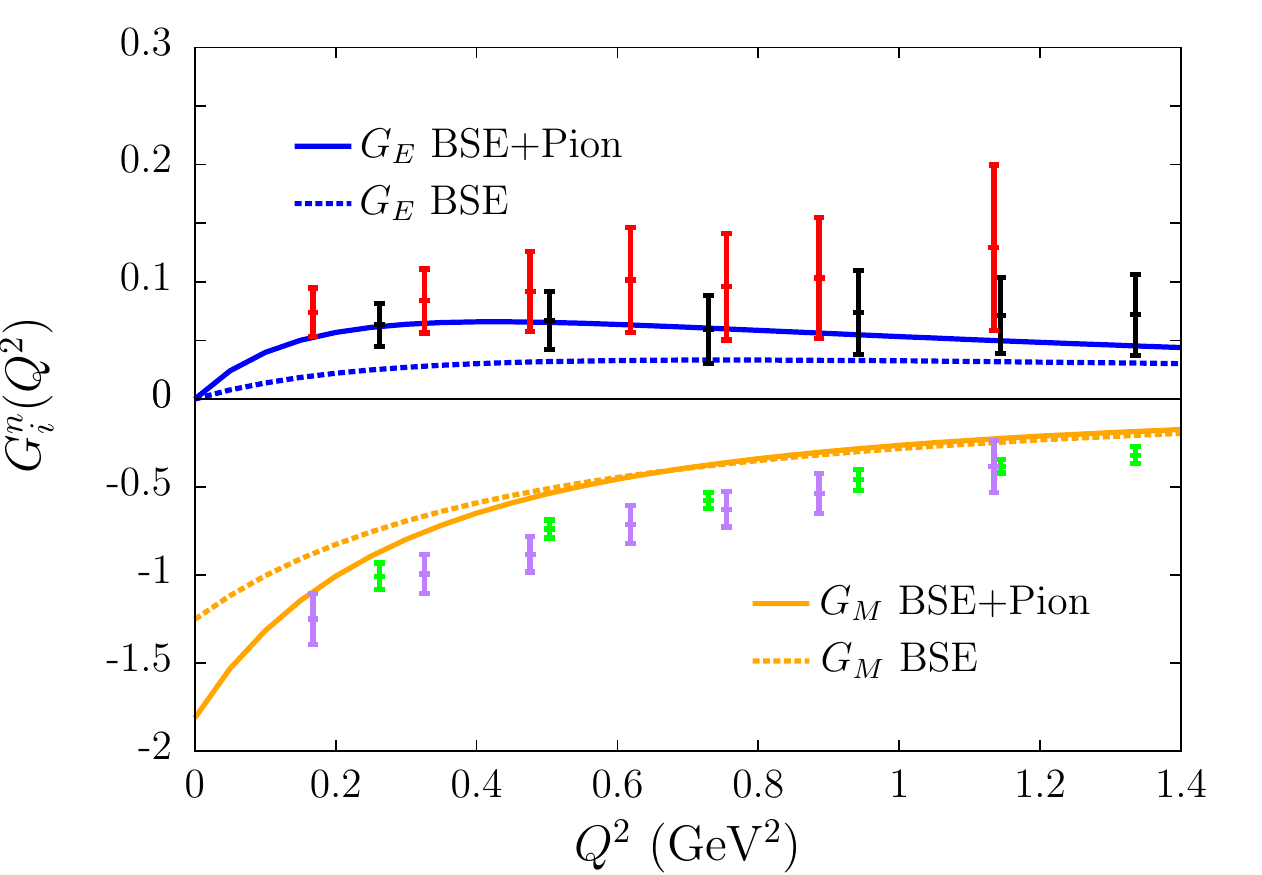}} \\
  \subfloat{\centering\includegraphics[width=1.0\columnwidth,clip=true,angle=0]{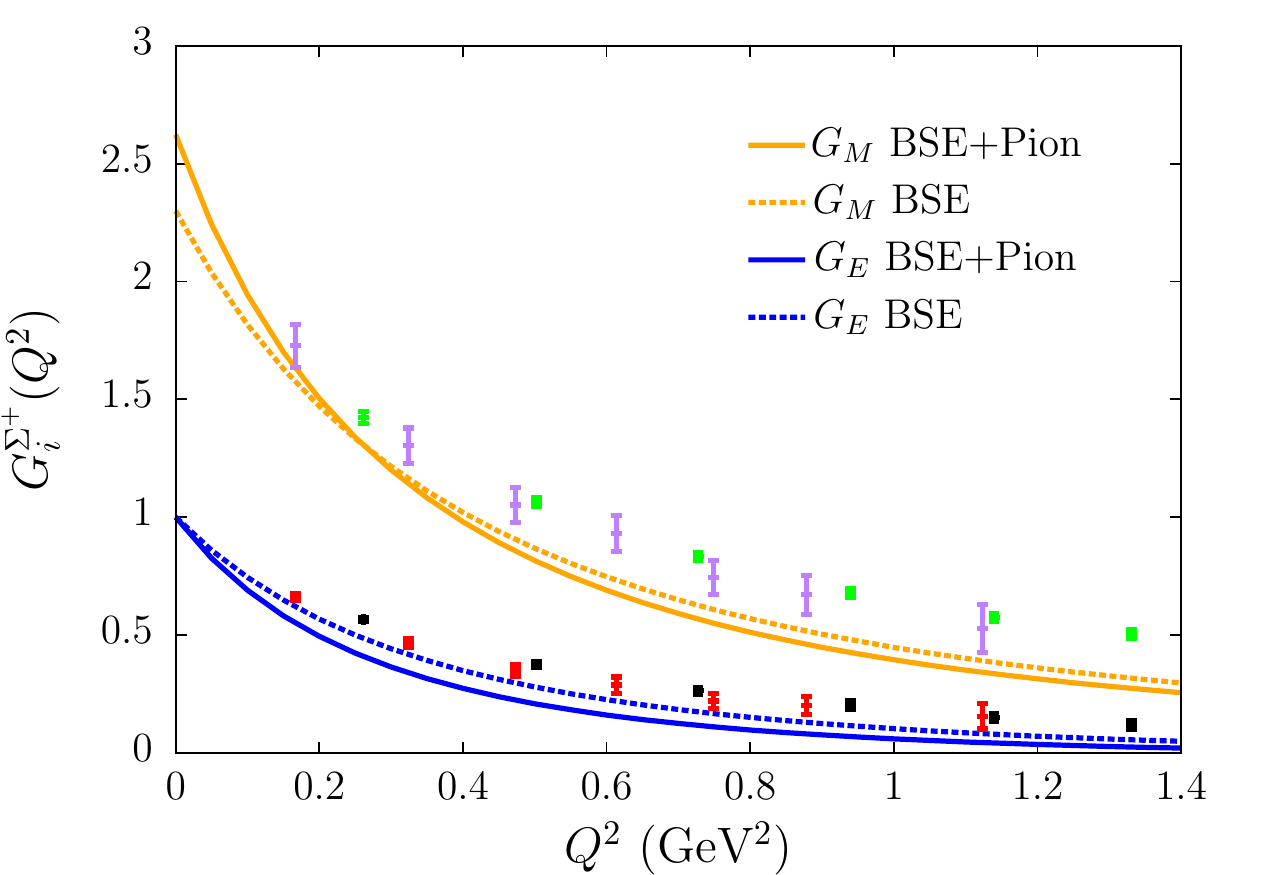}}
  \subfloat{\centering\includegraphics[width=1.0\columnwidth,clip=true,angle=0]{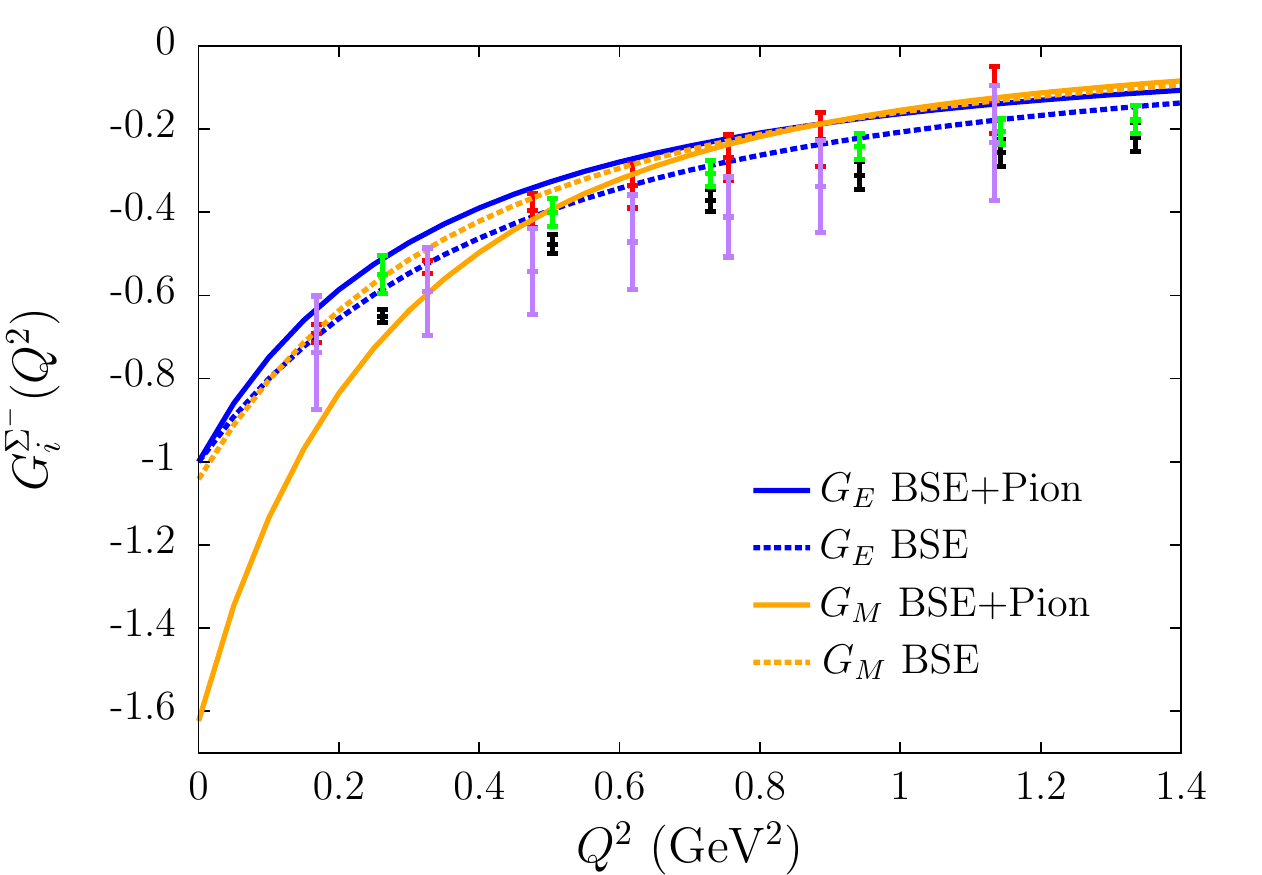}} \\
  \subfloat{\centering\includegraphics[width=1.0\columnwidth,clip=true,angle=0]{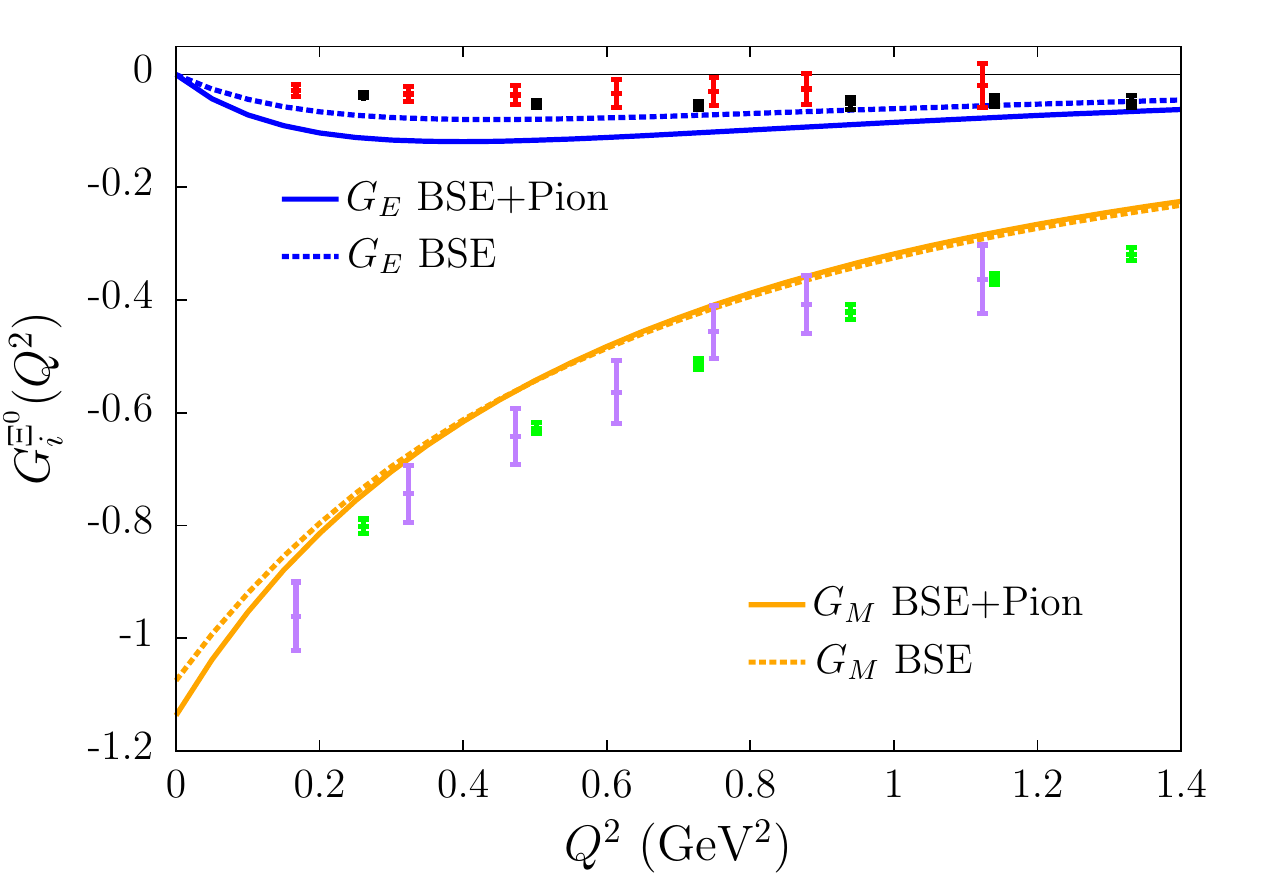}}
  \subfloat{\centering\includegraphics[width=1.0\columnwidth,clip=true,angle=0]{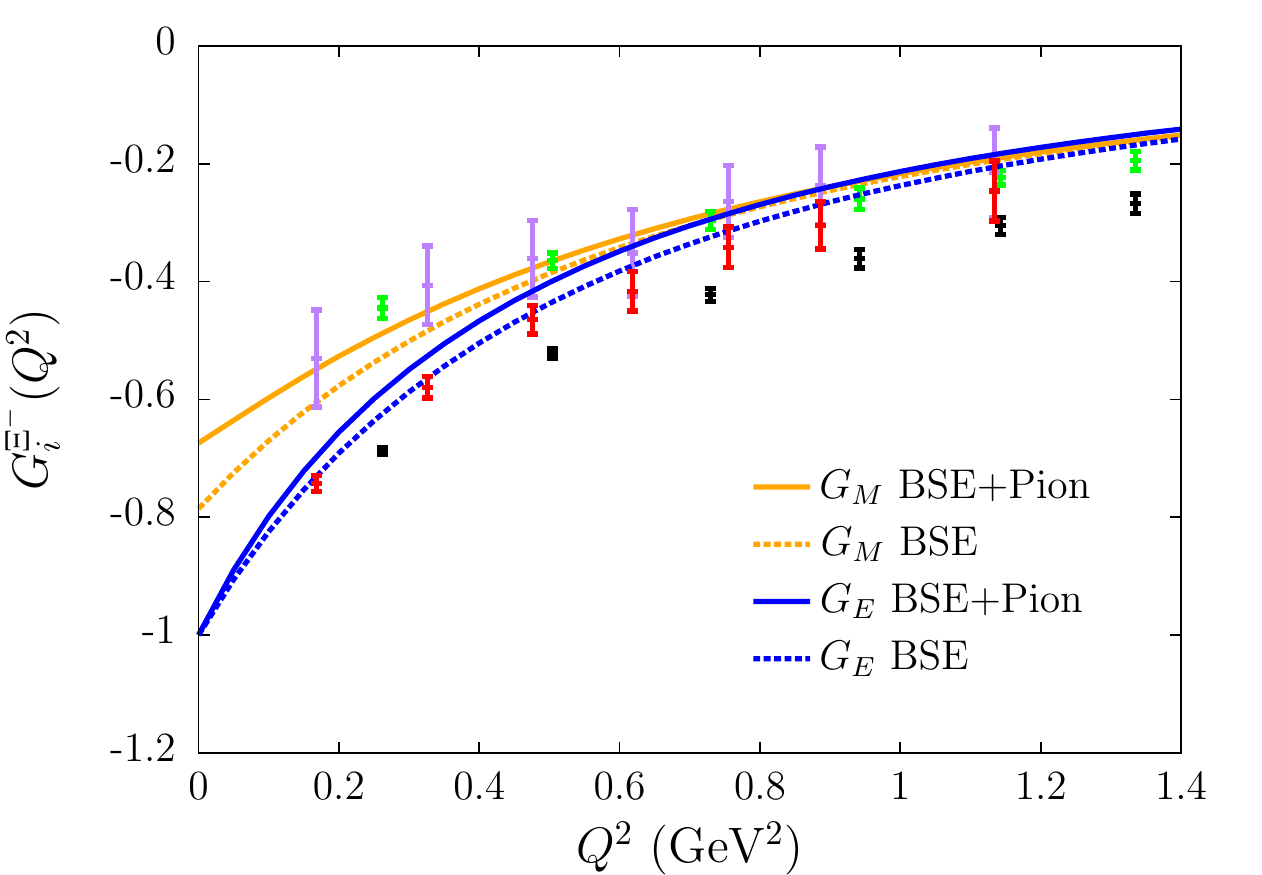}} \\
\caption{(Colour online) Electromagnetic form factors of the 
octet baryons with $i=(E,M)$ indicating
the Sachs electric and magnetic form factors. 
The plots show results from the vector meson dressing contributions
to the quark-photon vertex (BSE) and the case where the pion cloud also included (BSE+Pion). 
In all the plots 
the points with error bars correspond to the chiral extrapolation of lattice results presented
in Ref.~\cite{Shanahan:2014cga,Shanahan:2014uka}, which are based on two different lattice volumes. 
In each case the magnetic form factors are normalized such that the value at $Q^2=0$ represents 
the baryon magnetic moment in units of nuclear magnetons.}
\label{Fig:EMFFCOMB}
\end{figure*}
%===============================================================================

Our main results, presented in Fig.~\ref{Fig:EMFFCOMB}, compare the octet form 
factors calculated here with those obtained in Ref.~\cite{Shanahan:2014cga,Shanahan:2014uka} via
chiral extrapolation of lattice QCD simulations (on two different volumes) 
to the physical quark masses.
For the magnetic form factors of the neutron, $\Sigma^{-}$ and $\Xi^-$ 
it is evident the contribution from the pion cloud is significant, 
primarily at low $Q^2$. The effect of the pion
cloud on the nucleon electric form factors appears to improve 
the agreement with the lattice simulations (which are in quite good agreement 
with the empirical data), 
whereas the magnetic form factors are still underestimated.

The other members of the octet appear to have a similar behaviour.
However, for the $\Sigma^-$ the curvature
of $G_M$ is dramatically increased by the pion cloud. 
This behaviour matches the large increase
of the $\Sigma^-$ magnetic moment reported earlier. 
$G_E^{\Sigma^-}$ is consistent with the lattice results 
within the same level of accuracy found for the nucleon,
$\Sigma^+$ and $\Xi^0$. Finally, $G_M^{\Xi^-}$ shows  
outstanding agreement with the lattice data but
$G_E^{\Xi^-}$ is above the data, just as found for 
the other members of the octet. 

The possible explanation of these small, but 
non-negligible differences could well 
be a consequence of the fact that there are still systematic
errors from the lattice simulation and the chiral extrapolation technique.

%===============================================================================
\begin{table}[t!]
\addtolength{\tabcolsep}{3.8pt}
\addtolength{\extrarowheight}{3.0pt}
\begin{tabular}{c|cccccc}
\hline\hline
 & $r_E^{(PL)}$ & $r_E^{(BSE)}$ & $r_E$ & $r_M^{(PL)}$ & $r_M^{(BSE)}$ & $r_M$\\
\hline
proton   &  0.51 &  0.81 &  0.87   & 0.43 & 0.76 & 0.87   \\
neutron  & -0.19 & -0.20 & -0.37   & 0.39 & 0.74 & 0.91   \\
$\S^+$   &  0.53 &  0.85 &  0.96   & 0.45 & 0.76 & 0.88   \\
$\S^-$   &  0.46 &  0.74 &  0.86   & 0.48 & 0.80 & 0.96   \\
$\Xi^0$  &  0.17 &  0.37 &  0.49   & 0.35 & 0.62 & 0.66   \\
$\Xi^-$  &  0.44 &  0.69 &  0.76   & 0.42 & 0.62 & 0.51   \\
\hline\hline
\end{tabular}
\caption{Electric and magnetic radii (in fm). PL stands for a point-like quark,
BSE includes only the vector meson contributions to the dressed quark form factors, and
the final results include both BSE and the effect of the pion cloud.}
\label{Tab:Radii}
\end{table}
%===============================================================================

%===============================================================================
%===============================================================================
\section{CONCLUSIONS\label{sect:con}}
We have presented a study of the electromagnetic form factors of the baryon 
octet over an extended range of momentum transfer. The calculations were made
within the NJL model, using proper-time regularization to simulate confinement and
included dressing at the quark-photon vertices from vector mesons and
pion loops. This work was stimulated by the recent lattice QCD calculations of these quantities,
which presented results (after chiral extrapolation) at a discrete set of values
of $Q^2$ up to $1.4$ GeV$^2$. In comparing the model calculations with these lattice results
one must bear in mind that there may still be systematic errors at the level of 
10\% arising from lattice artifacts as well as the chiral extrapolation.

Overall, the level of agreement between the model calculations and the lattice results
is qualitatively impressive. 
We expect that the results presented here will stimulate calculations in other
approaches and trust that the comparison between those results, future lattice 
calculations and the results presented here will indeed lead to important new insights into
hadron structure.

%===============================================================================
%===============================================================================
\section*{ACKNOWLEDGEMENTS}
This material is based upon work supported by the U.S. Department of Energy, 
Office of Science, 
Office of Nuclear Physics, under contract number DE-AC02-06CH11357; 
the Australian Research 
Council through the ARC Centre of Excellence in Particle Physics at the Terascale, 
an ARC Australian Laureate Fellowship FL0992247 and DP151103101;
and the Grant in Aid for Scientific Research (Kakenhi) of the Japanese Ministry of
Education, Sports, Science and Technology, Project No. 25400270.

%===============================================================================
%===============================================================================

%

\begin{thebibliography}{99}
%
%\cite{Arrington:2006zm}
\bibitem{Arrington:2006zm} 
  J.~Arrington, C.~D.~Roberts and J.~M.~Zanotti,
  %``Nucleon electromagnetic form-factors,''
  J.\ Phys.\ G {\bf 34}, S23 (2007)
  [nucl-th/0611050].
  %%CITATION = NUCL-TH/0611050;%%
  %141 citations counted in INSPIRE as of 14 Jan 2015

%\cite{Brodsky:1973kr}
\bibitem{Brodsky:1973kr} 
  S.~J.~Brodsky and G.~R.~Farrar,
  %``Scaling Laws at Large Transverse Momentum,''
  Phys.\ Rev.\ Lett.\  {\bf 31}, 1153 (1973).
  %%CITATION = PRLTA,31,1153;%%
  %1554 citations counted in INSPIRE as of 01 Dec 2014

%\cite{Brodsky:1974vy}
\bibitem{Brodsky:1974vy} 
  S.~J.~Brodsky and G.~R.~Farrar,
  %``Scaling Laws for Large Momentum Transfer Processes,''
  Phys.\ Rev.\ D {\bf 11}, 1309 (1975).
  %%CITATION = PHRVA,D11,1309;%%
  %907 citations counted in INSPIRE as of 01 Dec 2014

%\cite{Farrar:1979aw}
\bibitem{Farrar:1979aw} 
  G.~R.~Farrar and D.~R.~Jackson,
  %``The Pion Form-Factor,''
  Phys.\ Rev.\ Lett.\  {\bf 43}, 246 (1979).
  %%CITATION = PRLTA,43,246;%%
  %344 citations counted in INSPIRE as of 01 Dec 2014

%\cite{Tadevosyan:2007yd}
\bibitem{Tadevosyan:2007yd} 
  V.~Tadevosyan {\it et al.}  [Jefferson Lab F(pi) Collaboration],
  %``Determination of the pion charge form-factor for Q**2 = 0.60-GeV**2 - 1.60-GeV**2,''
  Phys.\ Rev.\ C {\bf 75}, 055205 (2007)
  [nucl-ex/0607007].
  %%CITATION = NUCL-EX/0607007;%%
  %152 citations counted in INSPIRE as of 01 Dec 2014

%\cite{Jones:1999rz}
\bibitem{Jones:1999rz} 
  M.~K.~Jones {\it et al.}  [Jefferson Lab Hall A Collaboration],
  %``G(E(p)) / G(M(p)) ratio by polarization transfer in polarized e p ---> e polarized p,''
  Phys.\ Rev.\ Lett.\  {\bf 84}, 1398 (2000)
  [nucl-ex/9910005].
  %%CITATION = NUCL-EX/9910005;%%
  %722 citations counted in INSPIRE as of 01 Dec 2014

%\cite{Gayou:2001qt}
\bibitem{Gayou:2001qt} 
  O.~Gayou, K.~Wijesooriya, A.~Afanasev, M.~Amarian, K.~Aniol, S.~Becher, K.~Benslama and L.~Bimbot {\it et al.},
  %``Measurements of the elastic electromagnetic form-factor ratio mu(p) G(Ep) / G(Mp) via polarization transfer,''
  Phys.\ Rev.\ C {\bf 64}, 038202 (2001).
  %%CITATION = PHRVA,C64,038202;%%
  %213 citations counted in INSPIRE as of 01 Dec 2014

%\cite{Gayou:2001qd}
\bibitem{Gayou:2001qd} 
  O.~Gayou {\it et al.}  [Jefferson Lab Hall A Collaboration],
  %``Measurement of G(Ep) / G(Mp) in polarized-e p ---> e polarized-p to Q**2 = 5.6-GeV**2,''
  Phys.\ Rev.\ Lett.\  {\bf 88}, 092301 (2002)
  [nucl-ex/0111010].
  %%CITATION = NUCL-EX/0111010;%%
  %644 citations counted in INSPIRE as of 01 Dec 2014

%\cite{Ralston:2003mt}
\bibitem{Ralston:2003mt} 
  J.~P.~Ralston and P.~Jain,
  %``QCD form-factors and hadron helicity nonconservation,''
  Phys.\ Rev.\ D {\bf 69}, 053008 (2004)
  [hep-ph/0302043].
  %%CITATION = HEP-PH/0302043;%%
  %41 citations counted in INSPIRE as of 01 Dec 2014

%\cite{Kelly:2004hm}
\bibitem{Kelly:2004hm} 
  J.~J.~Kelly,
  %``Simple parametrization of nucleon form factors,''
  Phys.\ Rev.\ C {\bf 70}, 068202 (2004).
  %%CITATION = PHRVA,C70,068202;%%
  %233 citations counted in INSPIRE as of 01 Dec 2014

%\cite{Perdrisat:2006hj}
\bibitem{Perdrisat:2006hj} 
  C.~F.~Perdrisat, V.~Punjabi and M.~Vanderhaeghen,
  %``Nucleon Electromagnetic Form Factors,''
  Prog.\ Part.\ Nucl.\ Phys.\  {\bf 59}, 694 (2007)
  [hep-ph/0612014].
  %%CITATION = HEP-PH/0612014;%%
  %228 citations counted in INSPIRE as of 01 Dec 2014

%\cite{Bernauer:2010wm}
\bibitem{Bernauer:2010wm} 
  J.~C.~Bernauer {\it et al.}  [A1 Collaboration],
  %``High-precision determination of the electric and magnetic form factors of the proton,''
  Phys.\ Rev.\ Lett.\  {\bf 105}, 242001 (2010)
  [arXiv:1007.5076 [nucl-ex]].
  %%CITATION = ARXIV:1007.5076;%%
  %147 citations counted in INSPIRE as of 01 Dec 2014

%\cite{Puckett:2011xg}
\bibitem{Puckett:2011xg} 
  A.~J.~R.~Puckett, E.~J.~Brash, O.~Gayou, M.~K.~Jones, L.~Pentchev, C.~F.~Perdrisat, V.~Punjabi and K.~A.~Aniol {\it et al.},
  %``Final Analysis of Proton Form Factor Ratio Data at $\mathbf{Q^2 = 4.0}$, 4.8 and 5.6 GeV$\mathbf{^2}$,''
  Phys.\ Rev.\ C {\bf 85}, 045203 (2012)
  [arXiv:1102.5737 [nucl-ex]].
  %%CITATION = ARXIV:1102.5737;%%
  %53 citations counted in INSPIRE as of 01 Dec 2014

%\cite{Agashe:2014kda}
\bibitem{Agashe:2014kda} 
  K.~A.~Olive {\it et al.}  [Particle Data Group Collaboration],
  %``Review of Particle Physics,''
  Chin.\ Phys.\ C {\bf 38}, 090001 (2014).
  %%CITATION = CHPHD,C38,090001;%%
  %446 citations counted in INSPIRE as of 14 Jan 2015

%\cite{LeYaouanc:1976ne}
\bibitem{LeYaouanc:1976ne} 
  A.~Le Yaouanc, L.~Oliver, O.~Pene and J.~-C.~Raynal,
  %``Phenomenological SU(6) Breaking of Baryon Wave Functions and the Chromodynamic Spin Spin Force,''
  Phys.\ Rev.\ D {\bf 18}, 1591 (1978).
  %%CITATION = PHRVA,D18,1591;%%

%\cite{Isgur:1979be}
\bibitem{Isgur:1979be} 
  N.~Isgur and G.~Karl,
  %``Ground State Baryons in a Quark Model with Hyperfine Interactions,''
  Phys.\ Rev.\ D {\bf 20}, 1191 (1979).
  %%CITATION = PHRVA,D20,1191;%%

%\cite{Thomas:1981vc}
\bibitem{Thomas:1981vc} 
  A.~W.~Thomas, S.~Theberge and G.~A.~Miller,
  %``The Cloudy Bag Model of the Nucleon,''
  Phys.\ Rev.\ D {\bf 24}, 216 (1981).
  %%CITATION = PHRVA,D24,216;%%
  %358 citations counted in INSPIRE as of 01 Dec 2014

%\cite{Diakonov:1996sr}
\bibitem{Diakonov:1996sr} 
  D.~Diakonov, V.~Petrov, P.~Pobylitsa, M.~V.~Polyakov and C.~Weiss,
  %``Nucleon parton distributions at low normalization point in the large N(c) limit,''
  Nucl.\ Phys.\ B {\bf 480}, 341 (1996)
  [hep-ph/9606314].
  %%CITATION = HEP-PH/9606314;%%

%\cite{Diakonov:1997sj}
\bibitem{Diakonov:1997sj} 
  D.~Diakonov,
  %``Chiral quark - soliton model,''
  In *Peniscola 1997, Advanced school on non-perturbative quantum field physics* 1-55
  [hep-ph/9802298].
  %%CITATION = HEP-PH/9802298;%%

%\cite{Mineo:1999eq}
\bibitem{Mineo:1999eq} 
  H.~Mineo, W.~Bentz and K.~Yazaki,
  %``Quark distributions in the nucleon based on a relativistic three-body approach to the NJL model,''
  Phys.\ Rev.\ C {\bf 60}, 065201 (1999)
  [nucl-th/9907043].
  %%CITATION = NUCL-TH/9907043;%%
  %35 citations counted in INSPIRE as of 01 Dec 2014

%\cite{Cloet:2005pp}
\bibitem{Cloet:2005pp} 
  I.~C.~Clo\"et, W.~Bentz and A.~W.~Thomas,
  %``Nucleon quark distributions in a covariant quark-diquark model,''
  Phys.\ Lett.\ B {\bf 621}, 246 (2005)
  [hep-ph/0504229].
  %%CITATION = HEP-PH/0504229;%%
  %40 citations counted in INSPIRE as of 01 Jun 2014

%\cite{Cloet:2005rt}
\bibitem{Cloet:2005rt} 
  I.~C.~Clo\"et, W.~Bentz and A.~W.~Thomas,
  %``Spin-dependent structure functions in nuclear matter and the polarized EMC effect,''
  Phys.\ Rev.\ Lett.\  {\bf 95}, 052302 (2005)
  [nucl-th/0504019].
  %%CITATION = NUCL-TH/0504019;%%
  %43 citations counted in INSPIRE as of 01 Jun 2014

%\cite{Cloet:2007em}
\bibitem{Cloet:2007em} 
  I.~C.~Clo\"et, W.~Bentz and A.~W.~Thomas,
  %``Transversity quark distributions in a covariant quark-diquark model,''
  Phys.\ Lett.\ B {\bf 659}, 214 (2008)
  [arXiv:0708.3246 [hep-ph]].
  %%CITATION = ARXIV:0708.3246;%%
  %50 citations counted in INSPIRE as of 01 Jun 2014

%\cite{Bentz:2007zs}
\bibitem{Bentz:2007zs} 
  W.~Bentz, I.~C.~Clo\"et, T.~Ito, A.~W.~Thomas and K.~Yazaki,
  %``Polarized structure functions of nucleons and nuclei,''
  Prog.\ Part.\ Nucl.\ Phys.\  {\bf 61}, 238 (2008)
  [arXiv:0711.0392 [nucl-th]].
  %%CITATION = ARXIV:0711.0392;%%
  %10 citations counted in INSPIRE as of 01 Jun 2014

%\cite{Matevosyan:2011vj}
\bibitem{Matevosyan:2011vj} 
  H.~H.~Matevosyan, W.~Bentz, I.~C.~Cl\"oet and A.~W.~Thomas,
  %``Transverse Momentum Dependent Fragmentation and Quark Distribution Functions from the NJL-jet Model,''
  Phys.\ Rev.\ D {\bf 85}, 014021 (2012)
  [arXiv:1111.1740 [hep-ph]].
  %%CITATION = ARXIV:1111.1740;%%
  %25 citations counted in INSPIRE as of 14 Jan 2015

%\cite{Cloet:2014rja}
\bibitem{Cloet:2014rja} 
  I.~C.~Clo\"et, W.~Bentz and A.~W.~Thomas,
  %``Role of diquark correlations and the pion cloud in nucleon elastic form factors,''
  Phys.\ Rev.\ C {\bf 90}, no. 4, 045202 (2014)
  [arXiv:1405.5542 [nucl-th]].
  %%CITATION = ARXIV:1405.5542;%%
  %3 citations counted in INSPIRE as of 26 Jan 2015

%\cite{Aliev:2013jda}
\bibitem{Aliev:2013jda} 
  T.~M.~Aliev, K.~Azizi and M.~Savci,
  %``Electromagnetic form factors of octet baryons in QCD,''
  Phys.\ Lett.\ B {\bf 723}, 145 (2013)
  [arXiv:1303.6798 [hep-ph]].
  %%CITATION = ARXIV:1303.6798;%%
  %5 citations counted in INSPIRE as of 25 Nov 2014

%\cite{Cloet:2008re}
\bibitem{Cloet:2008re} 
  I.~C.~Clo\"et, G.~Eichmann, B.~El-Bennich, T.~Klahn and C.~D.~Roberts,
  %``Survey of nucleon electromagnetic form factors,''
  Few Body Syst.\  {\bf 46}, 1 (2009)
  doi:10.1007/s00601-009-0015-x
  [arXiv:0812.0416 [nucl-th]].
  %%CITATION = doi:10.1007/s00601-009-0015-x;%%
  %107 citations counted in INSPIRE as of 28 Feb 2016

%\cite{Eichmann:2011vu}
\bibitem{Eichmann:2011vu} 
  G.~Eichmann,
  %``Nucleon electromagnetic form factors from the covariant Faddeev equation,''
  Phys.\ Rev.\ D {\bf 84}, 014014 (2011)
  doi:10.1103/PhysRevD.84.014014
  [arXiv:1104.4505 [hep-ph]].
  %%CITATION = doi:10.1103/PhysRevD.84.014014;%%
  %76 citations counted in INSPIRE as of 28 Feb 2016

%\cite{Segovia:2014aza}
\bibitem{Segovia:2014aza} 
  J.~Segovia, I.~C.~Clo\"et, C.~D.~Roberts and S.~M.~Schmidt,
  %``Nucleon and $\Delta$ elastic and transition form factors,''
  Few Body Syst.\  {\bf 55}, 1185 (2014)
  doi:10.1007/s00601-014-0907-2, 10.1007/s00601-014-0908-1
  [arXiv:1408.2919 [nucl-th]].
  %%CITATION = doi:10.1007/s00601-014-0907-2, 10.1007/s00601-014-0908-1;%%
  %21 citations counted in INSPIRE as of 28 Feb 2016

%\cite{Lin:2007ap}
\bibitem{Lin:2007ap} 
  H.~-W.~Lin and K.~Orginos,
  %``First Calculation of Hyperon Axial Couplings from Lattice QCD,''
  Phys.\ Rev.\ D {\bf 79}, 034507 (2009)
  [arXiv:0712.1214 [hep-lat]].
  %%CITATION = ARXIV:0712.1214;%%

%\cite{Durr:2008zz}
\bibitem{Durr:2008zz} 
  S.~Durr, Z.~Fodor, J.~Frison, C.~Hoelbling, R.~Hoffmann, S.~D.~Katz, S.~Krieg and T.~Kurth {\it et al.},
  %``Ab-Initio Determination of Light Hadron Masses,''
  Science {\bf 322}, 1224 (2008)
  [arXiv:0906.3599 [hep-lat]].
  %%CITATION = ARXIV:0906.3599;%%
  %319 citations counted in INSPIRE as of 01 Dec 2014

%\cite{WalkerLoud:2008bp}
\bibitem{WalkerLoud:2008bp} 
  A.~Walker-Loud, H.~-W.~Lin, D.~G.~Richards, R.~G.~Edwards, M.~Engelhardt, G.~T.~Fleming, P.~.Hagler and B.~Musch {\it et al.},
  %``Light hadron spectroscopy using domain wall valence quarks on an Asqtad sea,''
  Phys.\ Rev.\ D {\bf 79}, 054502 (2009)
  [arXiv:0806.4549 [hep-lat]].
  %%CITATION = ARXIV:0806.4549;%%

%\cite{Aoki:2008sm}
\bibitem{Aoki:2008sm} 
  S.~Aoki {\it et al.}  [PACS-CS Collaboration],
  %``2+1 Flavor Lattice QCD toward the Physical Point,''
  Phys.\ Rev.\ D {\bf 79}, 034503 (2009)
  [arXiv:0807.1661 [hep-lat]].
  %%CITATION = ARXIV:0807.1661;%%

%\cite{Shanahan:2012wa}
\bibitem{Shanahan:2012wa} 
  P.~E.~Shanahan, A.~W.~Thomas and R.~D.~Young,
  %``Strong contribution to octet baryon mass splittings,''
  Phys.\ Lett.\ B {\bf 718}, 1148 (2013)
  [arXiv:1209.1892 [nucl-th]].
  %%CITATION = ARXIV:1209.1892;%%
  %11 citations counted in INSPIRE as of 25 Nov 2014

%\cite{Shanahan:2013apa}
\bibitem{Shanahan:2013apa} 
  P.~E.~Shanahan, A.~W.~Thomas, K.~Tsushima, R.~D.~Young and F.~Myhrer,
  %``Octet Spin Fractions and the Proton Spin Problem,''
  Phys.\ Rev.\ Lett.\  {\bf 110}, no. 20, 202001 (2013)
  [arXiv:1302.6300 [nucl-th]].
  %%CITATION = ARXIV:1302.6300;%%

%\cite{Borsanyi:2014jba}
\bibitem{Borsanyi:2014jba} 
  S.~Borsanyi {\it et al.},
  %``Ab initio calculation of the neutron-proton mass difference,''
  Science {\bf 347}, 1452 (2015)
  [arXiv:1406.4088 [hep-lat]].
  %%CITATION = ARXIV:1406.4088;%%
  %62 citations counted in INSPIRE as of 21 sept. 2015

%\cite{Shanahan:2014cga}
\bibitem{Shanahan:2014cga} 
  P.~E.~Shanahan, A.~W.~Thomas, R.~D.~Young, J.~M.~Zanotti, R.~Horsley, Y.~Nakamura, D.~Pleiter and P.~E.~L.~Rakow {\it et al.},
  %``Electric form factors of the octet baryons from lattice QCD and chiral extrapolation,''
  Phys.\ Rev.\ D {\bf 90}, 034502 (2014)
  [arXiv:1403.1965 [hep-lat]].
  %%CITATION = ARXIV:1403.1965;%%
  %5 citations counted in INSPIRE as of 29 Oct 2014

%\cite{Shanahan:2014uka}
\bibitem{Shanahan:2014uka} 
  P.~E.~Shanahan, A.~W.~Thomas, R.~D.~Young, J.~M.~Zanotti, R.~Horsley, Y.~Nakamura, D.~Pleiter and P.~E.~L.~Rakow {\it et al.},
  %``Magnetic form factors of the octet baryons from lattice QCD and chiral extrapolation,''
  Phys.\ Rev.\ D {\bf 89}, 074511 (2014)
  [arXiv:1401.5862 [hep-lat]].
  %%CITATION = ARXIV:1401.5862;%%

%\cite{Chodos:1974pn}
\bibitem{Chodos:1974pn} 
  A.~Chodos, R.~L.~Jaffe, K.~Johnson and C.~B.~Thorn,
  %``Baryon Structure in the Bag Theory,''
  Phys.\ Rev.\ D {\bf 10}, 2599 (1974).
  %%CITATION = PHRVA,D10,2599;%%

%\cite{Theberge:1980ye}
\bibitem{Theberge:1980ye} 
  S.~Theberge, A.~W.~Thomas and G.~A.~Miller,
  %``The Cloudy Bag Model. 1. The (3,3) Resonance,''
  Phys.\ Rev.\ D {\bf 22}, 2838 (1980)
  [Erratum-ibid.\ D {\bf 23}, 2106 (1981)].
  %%CITATION = PHRVA,D22,2838;%%

%\cite{Theberge:1981pu}
\bibitem{Theberge:1981pu} 
  S.~Theberge and A.~W.~Thomas,
  %``The Magnetic Moments of the Nucleon Octet Calculated in the Cloudy Bag Model,''
  Phys.\ Rev.\ D {\bf 25}, 284 (1982).
  %%CITATION = PHRVA,D25,284;%%

%\cite{Myhrer:1982sp}
\bibitem{Myhrer:1982sp} 
  F.~Myhrer and Z.~Xu,
  %``Baryon Masses In The Broken Chiral Quark Bag,''
  Phys.\ Lett.\ B {\bf 108}, 372 (1982).
  %%CITATION = PHLTA,B108,372;%%
  %10 citations counted in INSPIRE as of 01 Dec 2014

%\cite{Kubodera:1984qd}
\bibitem{Kubodera:1984qd} 
  K.~Kubodera, Y.~Kohyama, K.~Oikawa and C.~W.~Kim,
  %``Weak Interactions Form-factors of the Octet Baryons in the Cloudy Bag Model,''
  Nucl.\ Phys.\ A {\bf 439}, 695 (1985).
  %%CITATION = NUPHA,A439,695;%%
  %22 citations counted in INSPIRE as of 01 Jun 2014

%\cite{Tsushima:1988xv}
\bibitem{Tsushima:1988xv} 
  K.~Tsushima, T.~Yamaguchi, Y.~Kohyama and K.~Kubodera,
  %``Weak Interaction Form-factors and Magnetic Moments of Octet Baryons: Chiral Bag Model With Gluonic Effects,''
  Nucl.\ Phys.\ A {\bf 489}, 557 (1988).
  %%CITATION = NUPHA,A489,557;%%

%\cite{Yamaguchi:1989sx}
\bibitem{Yamaguchi:1989sx} 
  T.~Yamaguchi, K.~Tsushima, Y.~Kohyama and K.~Kubodera,
  %``Semileptonic Beta Decay Form-factors and Magnetic Moments of Octet Baryons: Recoil Effects and Center-of-mass Corrections in the Cloudy Bag Model Including Gluonic Effects,''
  Nucl.\ Phys.\ A {\bf 500}, 429 (1989).
  %%CITATION = NUPHA,A500,429;%%
  %20 citations counted in INSPIRE as of 25 Aug 2014

%\cite{Wagner:1998fi}
\bibitem{Wagner:1998fi} 
  G.~Wagner, A.~J.~Buchmann and A.~Faessler,
  %``Exchange currents in octet hyperon charge radii,''
  Phys.\ Rev.\ C {\bf 58}, 3666 (1998)
  [nucl-th/9809015].
  %%CITATION = NUCL-TH/9809015;%%

%\cite{Boros:1999tb}
\bibitem{Boros:1999tb} 
  C.~Boros and A.~W.~Thomas,
  %``Parton distributions for the octet and decuplet baryons,''
  Phys.\ Rev.\ D {\bf 60}, 074017 (1999)
  [hep-ph/9902372].
  %%CITATION = HEP-PH/9902372;%%

%\cite{Thomas:1999mu}
\bibitem{Thomas:1999mu} 
  A.~W.~Thomas and G.~Krein,
  %``Chiral corrections in hadron spectroscopy,''
  Phys.\ Lett.\ B {\bf 456}, 5 (1999)
  [nucl-th/9902013].
  %%CITATION = NUCL-TH/9902013;%%
  %20 citations counted in INSPIRE as of 01 Dec 2014

%\cite{Thomas:2000fa}
\bibitem{Thomas:2000fa} 
  A.~W.~Thomas and G.~Krein,
  %``Chiral aspects of hadron structure,''
  Phys.\ Lett.\ B {\bf 481}, 21 (2000)
  [nucl-th/0004008].
  %%CITATION = NUCL-TH/0004008;%%
  %7 citations counted in INSPIRE as of 01 Dec 2014

%\cite{Bass:2009ed}
\bibitem{Bass:2009ed} 
  S.~D.~Bass and A.~W.~Thomas,
  %``The Nucleon's octet axial-charge g(A)**(8) with chiral corrections,''
  Phys.\ Lett.\ B {\bf 684}, 216 (2010)
  [arXiv:0912.1765 [hep-ph]].
  %%CITATION = ARXIV:0912.1765;%%

%\cite{Silva:2005vp}
\bibitem{Silva:2005vp} 
  A.~Silva, D.~Urbano and K.~Goeke,
  %``Baryon form factors in the chiral quark-soliton model,''
  Nucl.\ Phys.\ A {\bf 755}, 290 (2005).
  %%CITATION = NUPHA,A755,290;%%
  %1 citations counted in INSPIRE as of 01 Dec 2014

%\cite{Ramalho:2012pu}
\bibitem{Ramalho:2012pu} 
  G.~Ramalho, K.~Tsushima and A.~W.~Thomas,
  %``Octet Baryon Electromagnetic form Factors in Nuclear Medium,''
  J.\ Phys.\ G {\bf 40}, 015102 (2013)
  [arXiv:1206.2207 [hep-ph]].
  %%CITATION = ARXIV:1206.2207;%%
  %12 citations counted in INSPIRE as of 25 Nov 2014

%\cite{Carrillo-Serrano:2014zta}
\bibitem{Carrillo-Serrano:2014zta} 
  M.~E.~Carrillo-Serrano, I.~C.~Clo\"et and A.~W.~Thomas,
  %``SU(3)-flavor breaking in octet baryon masses and axial couplings,''
  Phys.\ Rev.\ C {\bf 90}, no. 6, 064316 (2014)
  [arXiv:1409.1653 [nucl-th]].
  %%CITATION = ARXIV:1409.1653;%%

%\cite{Sanchis-Alepuz:2015fcg}
\bibitem{Sanchis-Alepuz:2015fcg} 
  H.~Sanchis-Alepuz and C.~S.~Fischer,
  %``Hyperon elastic electromagnetic form factors in the space-like momentum region,''
  Eur.\ Phys.\ J.\ A {\bf 52}, no. 2, 34 (2016)
  doi:10.1140/epja/i2016-16034-1
  [arXiv:1512.00833 [hep-ph]].
  %%CITATION = doi:10.1140/epja/i2016-16034-1;%%
  %2 citations counted in INSPIRE as of 08 Mar 2016

%\cite{Young:2009zb}
\bibitem{Young:2009zb} 
  R.~D.~Young and A.~W.~Thomas,
  %``Octet baryon masses and sigma terms from an SU(3) chiral extrapolation,''
  Phys.\ Rev.\ D {\bf 81}, 014503 (2010)
  [arXiv:0901.3310 [hep-lat]].
  %%CITATION = ARXIV:0901.3310;%%

%\cite{Nambu:1961tp}
\bibitem{Nambu:1961tp} 
  Y.~Nambu and G.~Jona-Lasinio,
  %``Dynamical Model of Elementary Particles Based on an Analogy with Superconductivity. 1.,''
  Phys.\ Rev.\  {\bf 122}, 345 (1961).
  %%CITATION = PHRVA,122,345;%%
  %3936 citations counted in INSPIRE as of 21 May 2014

%\cite{Nambu:1961fr}
\bibitem{Nambu:1961fr} 
  Y.~Nambu and G.~Jona-Lasinio,
  %``Dynamical Model Of Elementary Particles Based On An Analogy With Superconductivity. Ii,''
  Phys.\ Rev.\  {\bf 124}, 246 (1961).
  %%CITATION = PHRVA,124,246;%%
  %1883 citations counted in INSPIRE as of 21 May 2014

%\cite{Carrillo-Serrano:2015uca}
\bibitem{Carrillo-Serrano:2015uca} 
  M.~E.~Carrillo-Serrano, W.~Bentz, I.~C.~Clo\"et and A.~W.~Thomas,
  %``$\rho$ meson form factors in a confining Nambu–Jona-Lasinio model,''
  Phys.\ Rev.\ C {\bf 92}, no. 1, 015212 (2015)
  [arXiv:1504.08119 [nucl-th]].
  %%CITATION = ARXIV:1504.08119;%%

%\cite{Ninomiya:2014kja}
\bibitem{Ninomiya:2014kja} 
  Y.~Ninomiya, W.~Bentz and I.~C.~Clo\"et,
  %``Dressed Quark Mass Dependence of Pion and Kaon Form Factors,''
  Phys.\ Rev.\ C {\bf 91}, no. 2, 025202 (2015)
  [arXiv:1406.7212 [nucl-th]].
  %%CITATION = ARXIV:1406.7212;%%
  %1 citations counted in INSPIRE as of 24 Apr 2015

%\cite{Klevansky:1992qe}
\bibitem{Klevansky:1992qe} 
  S.~P.~Klevansky,
  %``The Nambu-Jona-Lasinio model of quantum chromodynamics,''
  Rev.\ Mod.\ Phys.\  {\bf 64}, 649 (1992).
  %%CITATION = RMPHA,64,649;%%
  %1010 citations counted in INSPIRE as of 21 May 2014

%\cite{Hatsuda:1994pi}
\bibitem{Hatsuda:1994pi} 
  T.~Hatsuda and T.~Kunihiro,
  %``QCD phenomenology based on a chiral effective Lagrangian,''
  Phys.\ Rept.\  {\bf 247}, 221 (1994)
  [hep-ph/9401310].
  %%CITATION = HEP-PH/9401310;%%
  %1090 citations counted in INSPIRE as of 21 May 2014

%\cite{Ebert:1996vx}
\bibitem{Ebert:1996vx} 
  D.~Ebert, T.~Feldmann and H.~Reinhardt,
  %``Extended NJL model for light and heavy mesons without q - anti-q thresholds,''
  Phys.\ Lett.\ B {\bf 388}, 154 (1996)
  [hep-ph/9608223].
  %%CITATION = HEP-PH/9608223;%%
  %107 citations counted in INSPIRE as of 21 May 2014

%\cite{Hellstern:1997nv}
\bibitem{Hellstern:1997nv} 
  G.~Hellstern, R.~Alkofer and H.~Reinhardt,
  %``Diquark confinement in an extended NJL model,''
  Nucl.\ Phys.\ A {\bf 625}, 697 (1997)
  [hep-ph/9706551].
  %%CITATION = HEP-PH/9706551;%%
  %60 citations counted in INSPIRE as of 21 May 2014

%\cite{Bentz:2001vc}
\bibitem{Bentz:2001vc} 
  W.~Bentz and A.~W.~Thomas,
  %``The Stability of nuclear matter in the Nambu-Jona-Lasinio model,''
  Nucl.\ Phys.\ A {\bf 696}, 138 (2001)
  [nucl-th/0105022].
  %%CITATION = NUCL-TH/0105022;%%
  %92 citations counted in INSPIRE as of 21 May 2014

%\cite{Afnan:1977pi}
\bibitem{Afnan:1977pi} 
  I.~R.~Afnan and A.~W.~Thomas,
  %``Fundamentals of Three-Body Scattering Theory,''
  Top.\ Curr.\ Phys.\  {\bf 2}, 1 (1977).
  %%CITATION = WSSED,2,1;%%

%\cite{Vogl:1991qt}
\bibitem{Vogl:1991qt} 
  U.~Vogl and W.~Weise,
  %``The Nambu and Jona Lasinio model: Its implications for hadrons and nuclei,''
  Prog.\ Part.\ Nucl.\ Phys.\  {\bf 27}, 195 (1991).
  %%CITATION = PPNPD,27,195;%%
  %590 citations counted in INSPIRE as of 21 May 2014

%\cite{Ishii:1995bu}
\bibitem{Ishii:1995bu} 
  N.~Ishii, W.~Bentz and K.~Yazaki,
  %``Baryons in the NJL model as solutions of the relativistic Faddeev equation,''
  Nucl.\ Phys.\ A {\bf 587}, 617 (1995).
  %%CITATION = NUPHA,A587,617;%%
  %69 citations counted in INSPIRE as of 21 May 2014

%\cite{Ishii:1993np}
\bibitem{Ishii:1993np} 
  N.~Ishii, W.~Bentz and K.~Yazaki,
  %``Faddeev approach to the nucleon in the Nambu-Jona-Lasinio (NJL) model,''
  Phys.\ Lett.\ B {\bf 301}, 165 (1993).
  %%CITATION = PHLTA,B301,165;%%
  %60 citations counted in INSPIRE as of 01 Jun 2014

%\cite{Ishii:1993rt}
\bibitem{Ishii:1993rt} 
  N.~Ishii, W.~Bentz and K.~Yazaki,
  %``Solution of the relativistic three quark Faddeev equation in the Nambu-Jona-Lasinio (NJL) model,''
  Phys.\ Lett.\ B {\bf 318}, 26 (1993).
  %%CITATION = PHLTA,B318,26;%%
  %57 citations counted in INSPIRE as of 01 Jun 2014

%\cite{Buck:1992wz}
\bibitem{Buck:1992wz}
  A.~Buck, R.~Alkofer and H.~Reinhardt,
  %``Baryons as bound states of diquarks and quarks in the Nambu-Jona-Lasinio model,''
  Phys.\ Lett.\ B {\bf 286}, 29 (1992).
  %%CITATION = PHLTA,B286,29;%%
  %105 citations counted in INSPIRE as of 21 May 2014

\end{thebibliography}
\end{document}